\documentclass[aps,prl,reprint,groupedaddress]{revtex4-1}
\usepackage[pdftex]{graphicx}
\usepackage{epstopdf}
\usepackage{amsmath}

\begin{document}


\title{Vibrationally coherent crossing and coupling of electronic states during internal conversion in $\beta$-carotene}


\author{M. Liebel}
\author{C. Schnedermann}
\author{P. Kukura}
\email{philipp.kukura@chem.ox.ac.uk}
\affiliation{Physical and Theoretical Chemistry Laboratory, South Parks Road, Oxford OX1 3QZ, UK}


\date{\today}

\begin{abstract}
Coupling of nuclear and electronic degrees of freedom mediates energy flow in molecules after optical excitation. The associated coherent dynamics in polyatomic systems, however, remain experimentally unexplored. Here, we combined transient absorption spectroscopy with electronic population control to reveal nuclear wavepacket dynamics during the S$_2$ $\rightarrow$ S$_1$ internal conversion in  $\beta$-carotene. We show that passage through a conical intersection is vibrationally coherent and thereby provides direct feedback on the role of different vibrational coordinates in the breakdown of the Born-Oppenheimer approximation.
\end{abstract}


\maketitle

Conical intersections (CI) between potential energy surfaces are a dominant concept in theoretical treatments of ultrafast electronic 
dynamics in polyatomic molecules \cite{Teller1937,Yarkony1996,Levine2007,Domcke2012}. According to this description, displacement along a subset of molecular degrees of freedom mediates efficient coupling of  electronic states, but experimental information on the identity of these coordinates is lacking. Vibrational coherence (VC) \cite{Ruhman1988,Rose1988,Fragnito1989,Zhu1994,Ashworth1996,Banin1994,Fujiyoshi2003} is an ideal tool to study the structural aspects of CI dynamics, because the evolution of nuclear wavepackets after electronic surface crossing has been predicted to depend strongly on how the respective vibrational coordinate is involved in the CI \cite{Koppel1984,Kuhl2002}. Early ultrafast studies of diatomic molecules have directly revealed coherent vibronic coupling \cite{Rose1988}, but for polyatomic molecules, VC after surface crossing has only been reported in low-frequency coordinates. Vibrational periods comparable to or longer than the surface crossing time \cite{Qing1994,Seel1997}, however, make it difficult to distinguish whether the VC has survived passage through the CI or was generated by the surface crossing. Therefore, no experimental information exists on how VCs are affected by passage through a CI and whether vibrationally coherent coupling of electronic states occurs at all in polyatomics.

The lack of experimental studies using VC as a sensitive probe of CI dynamics is largely a consequence of technical difficulties. Time-domain observation of VC requires high temporal resolution ($\textless$ 15 fs) to access full vibrational spectra ($\textless$ 2000 $\mathrm{cm}^{-1}$) \cite{Lanzani2001} and  broad probing bandwidths to simultaneously monitor the electronic states involved in the CI. In addition, spectroscopic signatures of VCs in the time-domain are orders of magnitude smaller than their electronic counterparts demanding high spectroscopic sensitivity. 
Even under optimized experimental conditions \cite{Kobayashi2001}, however, the isolation of excited state signatures is challenging because of the unavoidable and often dominant presence of ground state and solvent VCs \cite{Kraack2011}. As a result, no clear observation of multiple excited state nuclear wavepackets generated directly by the excitation pulse has been reported to date. In this work, we used highly time-resolved pump-probe (PP) spectroscopy combined with electronic population control to generate and monitor the fate of excited state VCs during the $\mathrm{S}_{2} \rightarrow \mathrm{S}_{1}$ internal conversion (IC) in $\beta$-carotene in solution.

\begin{figure}[ht]
\includegraphics{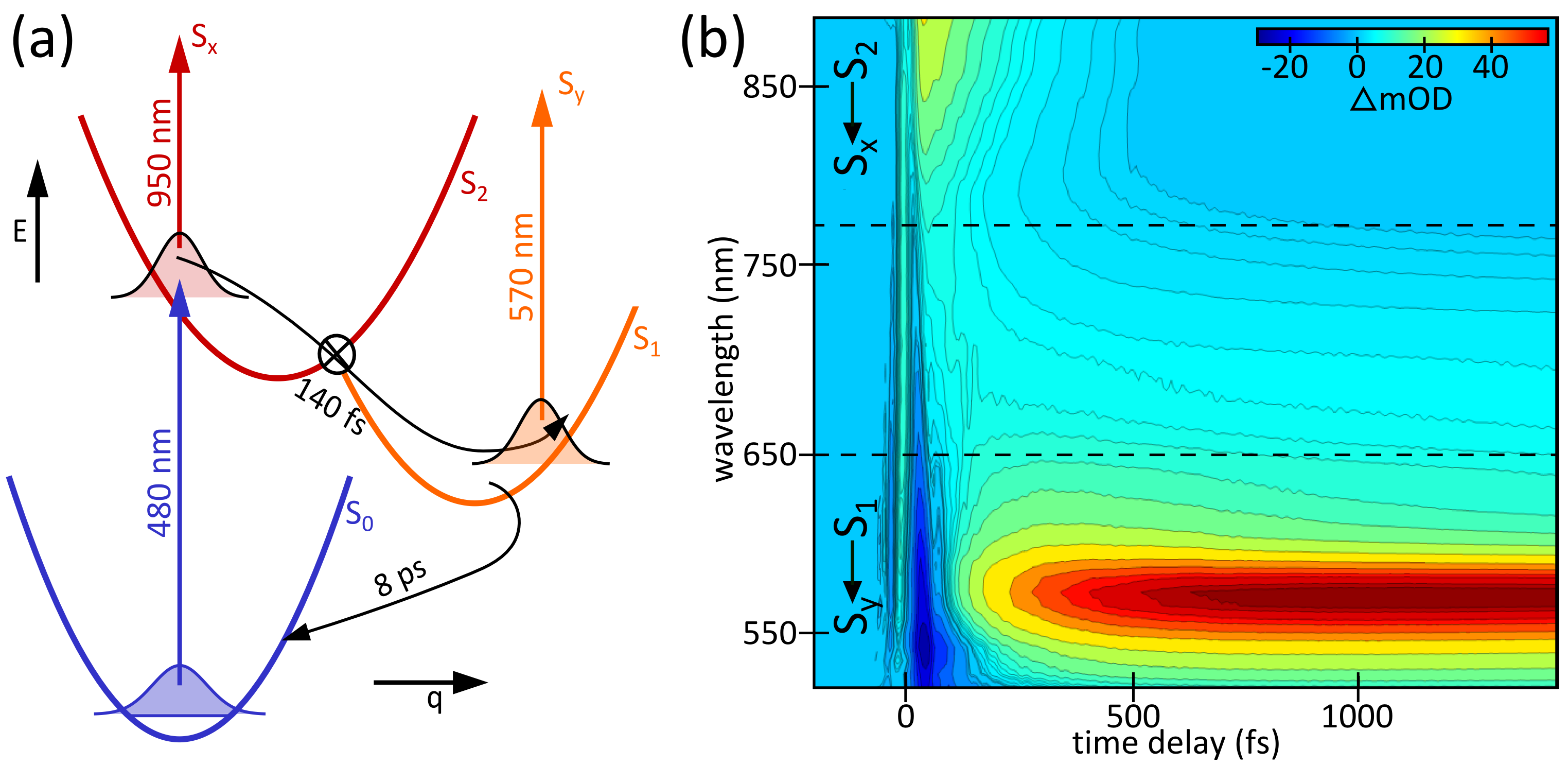}
\caption{\label{fig:figA} Photophysics and electronic dynamics of $\beta$-carotene. (a) Energy level scheme for $\beta$-carotene with the absorption maxima of the three lowest electronic singlet states. (b) Chirp corrected differential absorbance map for $\beta$-carotene in toluene. The sample was flown through a sample cell at OD = 0.7/200 $\mu\mathrm{m}$. Pump and probe intensities were adjusted to be 71 and 7 $\mathrm{mJ}\cdot\mathrm{cm}^{-2}$ at 75 and 50 $\mu$m beam diameters, respectively.}
\end{figure}

Photoexcitation of $\beta$-carotene in toluene populates the second excited singlet state S$_{2}$ that decays with a 140 fs time constant into S$_{1}$ (FIG. \ref{fig:figA}a) \cite{Macpherson1998}. The three lowest singlet states \cite{PerezLustres2007} are easily identified by their well separated electronic transitions, with absorption maxima at 950, 570 and 480 nm. To monitor the coherent vibronic dynamics of $\beta$-carotene, we performed a PP experiment with a 12 fs pump centered at 480 nm and a $\sim$300 fs chirped probe pulse (520 - 900 nm), a combination that allows for overal $\textless$ 15 fs time-resolution with spectrally resolved broadband probing \cite{Liebel2013}. One-photon absorption to S$_{1}$ is symmetry-forbidden ensuring that our signals are dominated by the dynamics following $\mathrm{S}_{2}$ $\leftarrow$ $\mathrm{S}_{0}$ excitation. The $\mathrm{S}_{2}$ $\rightarrow$ $\mathrm{S}_{1}$ IC is fast compared to common vibrational dephasing times so that most of the VC created by photoexcitation does not dephase on the timescale of IC. 

FIG. \ref{fig:figA}b depicts the wavelength-dependent differential absorbance of $\beta$-carotene as a function of pump-probe delay. Early time delays ($\textless$ 100 fs) exhibit broadband signatures of the coherent artifact \cite{Kovalenko1999}, a weak bleach at wavelengths $\textless$530 nm, signs of stimulated emission in the 530 - 650 nm window \cite{Andersson1995} and excited state absorption (ESA) from $\mathrm{S}_{2}$ in the near-infrared (NIR, $\textgreater$800 nm). The decay of the $\mathrm{S}_{x}$ $\leftarrow$ $\mathrm{S}_{2}$ and the rise of the $\mathrm{S}_{y}$ $\leftarrow$ $\mathrm{S}_{1}$ ESAs proceeds with the same time-constant (140 fs) in agreement with previous results \cite{Macpherson1998}.

\begin{figure}[ht]
\includegraphics{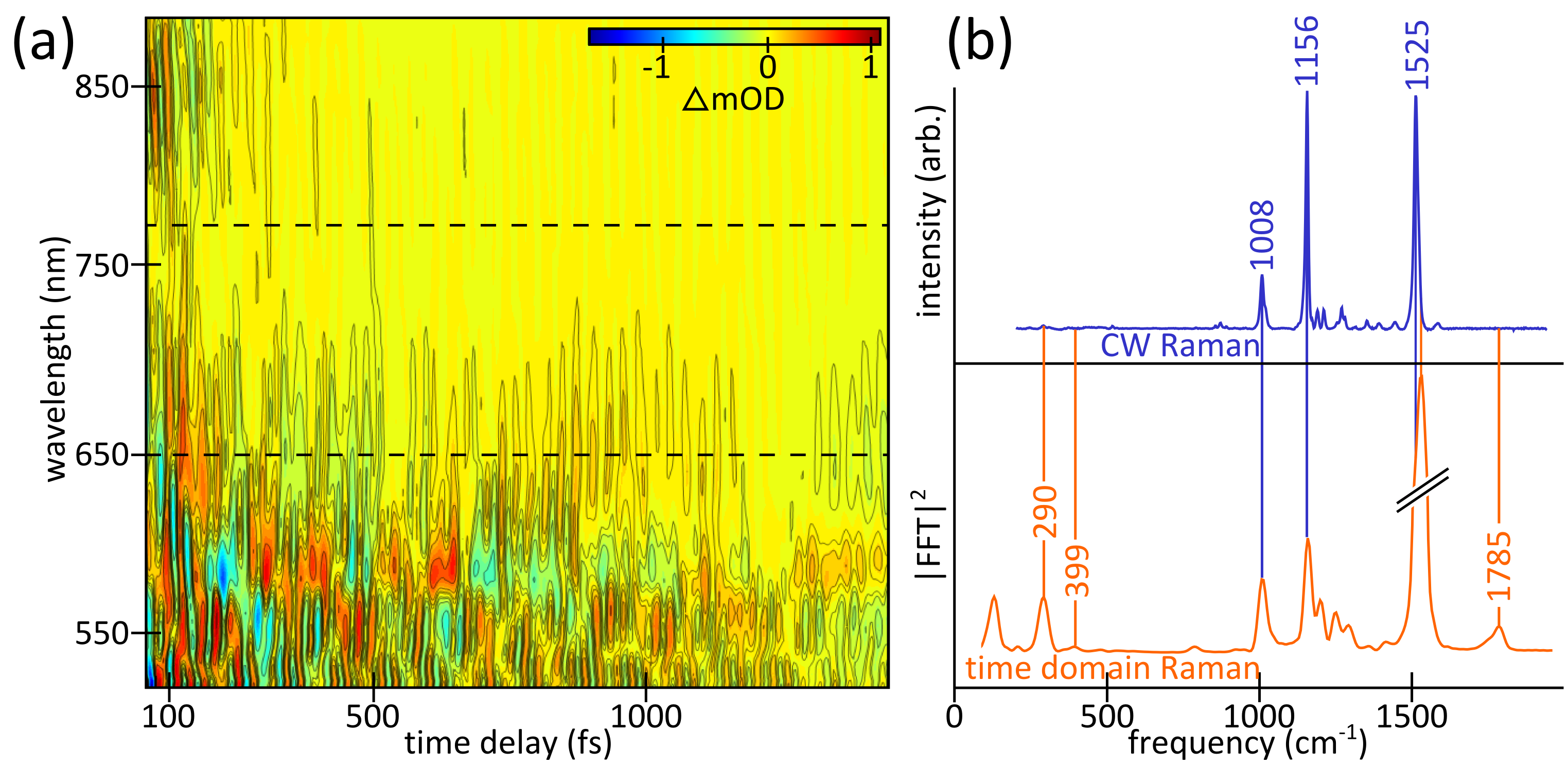}
\caption{\label{fig:figB} $\beta$-carotene vibrational coherence and Raman spectra. (a) Residual vibrational coherence after removal of the electronic kinetics obtained by global fitting. The black contour lines are set at $\pm$ 450, $\pm$ 200, $\pm$ 100 and $\pm$ 65 $\mu$OD to highlight regions of predominant coherent activity. (b) Averaged Fourier power spectrum in the 540 - 650 nm spectral region (orange) in comparison with a CW Raman spectrum of $\beta$-carotene (blue). $\mathrm{S}_{1}$ bands are indicated by orange, $\mathrm{S}_{0}$ bands by blue lines, respectively.
 }
\end{figure}

We truncated the data at $\Delta$t = 85 fs to avoid contamination by the coherent artifact before globally fitting the remaining differential absorbance. The coherence map after subtraction of the slowly varying electronic kinetics \cite{Liebel2013} exhibits three major regions of activity as a function of wavelength (FIG. \ref{fig:figB}a). The NIR coherence ($\textgreater$800 nm, $\mathrm{S}_{2}$ ESA) decays rapidly with a time constant comparable to the  $\mathrm{S}_{2}$ lifetime, VC in the visible (540 - 650 nm, $\mathrm{S}_{1}$ ESA) dephases more slowly and exhibits a phase jump around 570 nm \cite{Kumar2001,Kobayashi2008}, while weak and slowly decaying oscillations separate the two ESAs (650 - 800 nm). Comparison with  FIG. \ref{fig:figA}b suggests that the NIR and visible regions are dominated by $\mathrm{S}_{2}$ and $\mathrm{S}_{1}$ VC, respectively, while the transition region (650 - 800 nm) highlights the fact that $\mathrm{S}_{0}$ and solvent coherences contribute over the full observation window. 

To gain further insight into the spectral content of the VCs, we performed a Fourier transform and averaged over the 540 - 650 nm spectral window to capture a representative spectrum of the coherent activity in the $\mathrm{S}_{1}$ region. In addition to strong $\mathrm{S}_{0}$ signatures at 1008 and 1156 $\mathrm{cm}^{-1}$, clear $\mathrm{S}_{1}$ marker bands emerge, such as the well-established 1785 $\mathrm{cm}^{-1}$ \cite{Hashimoto1989,Macpherson1998} and the recently reported 290 and 399 $\mathrm{cm}^{-1}$ bands (FIG. \ref{fig:figB}b) \cite{Liebel2013,Kraack2013}. Although these spectra suggest that VC in multiple degrees of freedom is transferred through the CI from $\mathrm{S}_{2}$ into $\mathrm{S}_{1}$, the additional presence of solvent and $\mathrm{S}_{0}$ coherences greatly diminishes the visibility of individual $\mathrm{S}_{1}$ coherences. 

\begin{figure}[ht]
\includegraphics{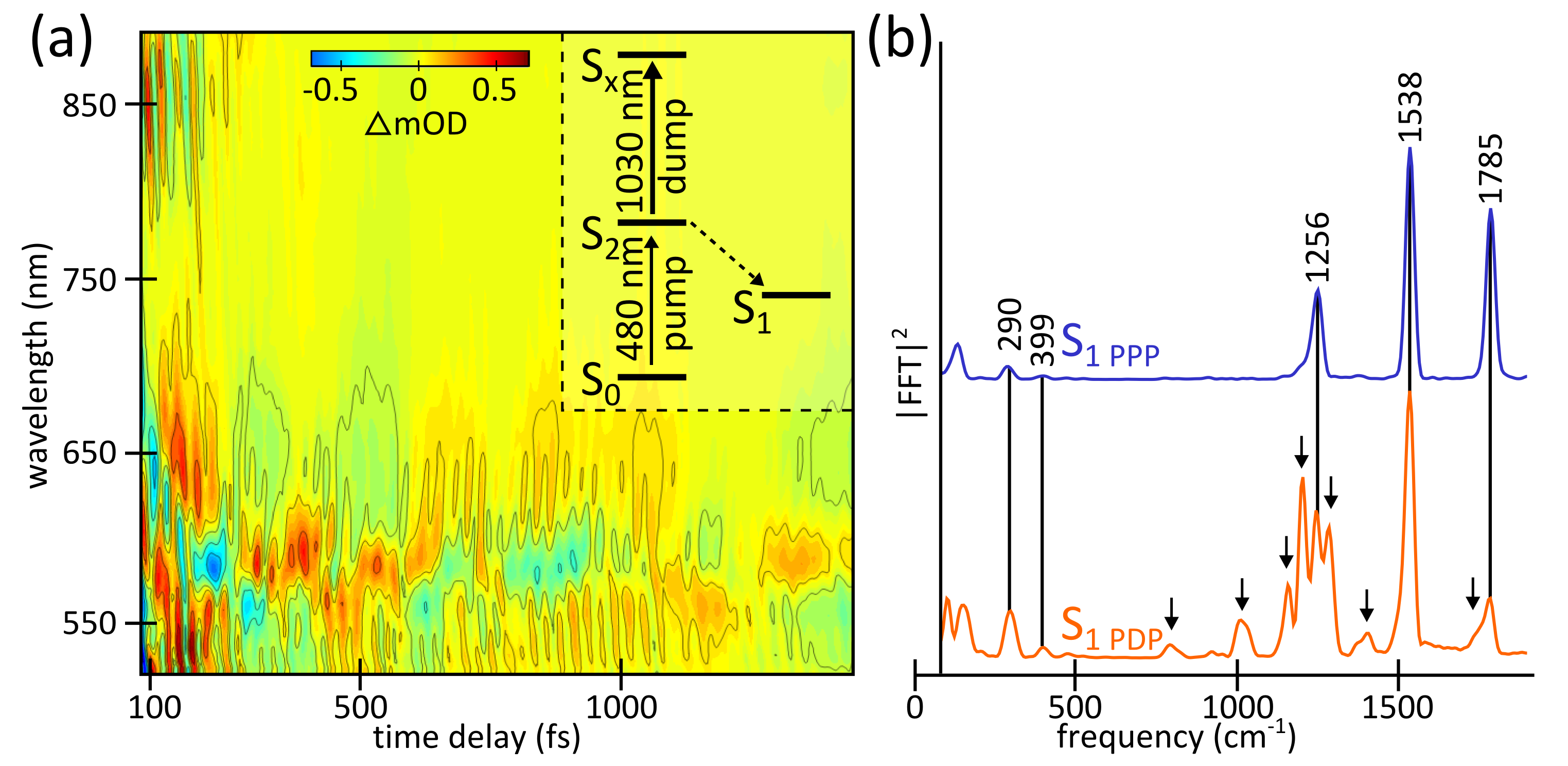}
\caption{\label{fig:figC} Electronic population control isolates excited state vibrational coherence. (a) $\mathrm{VC}_{OmO}$ coherence map, only coherence from $\mathrm{S}_{2}$ and $\mathrm{S}_{1}$ is present (compare FIG. \ref{fig:figB}a). The black contour lines are set at $\pm$ 500, $\pm$ 300 and $\pm$ 65 $\mu\mathrm{OD}$ to highlight regions of predominant coherent activity. A dump pulse centered at 1030 nm, 1 ps duration with an intensity of 150  $\mathrm{mJ}\cdot\mathrm{cm}^{-2}$  was used. (b) Comparison of a Fourier power spectrum obtained by broadband impulsive vibrational spectroscopy (blue) to the $\mathrm{S}_{1}$ spectrum obtained by IC from $\mathrm{S}_{2}$ (orange), the spectral intensities were corrected for time resolution (15 and 14 fs respectively). Common bands are indicated by black lines, enhanced bands by arrows.
 }
\end{figure}

We therefore designed an experimental approach that allows us to isolate VCs that originate from a specific electronic state of interest. Extending the two pulse PP experiment by a third “dump” pulse \cite{Weigel2011}, time coincident with the pump and resonant with the $\mathrm{S}_{x}$ $\leftarrow$ $\mathrm{S}_{2}$ transition, selectively removes $\mathrm{S}_{2}$ population (FIG. \ref{fig:figC}a inset). The duration of the dump pulse was chosen in such a way that it cannot impulsively create VCs with frequencies $\textgreater$50 $\mathrm{cm}^{-1}$. By comparing the PP to the pump-dump-probe (PDP) experiment after subtraction of the exponential electronic kinetics, we were able to isolate $\mathrm{S}_{2}$ VC as outlined in Eqs. \ref{eq:eqA}a-c

\begin{subequations}\label{eq:eqA}
\begin{align}
&\mathrm{VC}_{OFF} =\chi_{\beta car_{S_{2}}} + \chi_{\beta car_{S_{0}}} + \chi_{solvent}\\
&\mathrm{VC}_{ON} =(1 - x)\cdot\chi_{\beta car_{S_{2}}} + \chi_{\beta car_{S_{0}}} + \chi_{solvent}\\
&\mathrm{VC}_{OmO} = \mathrm{VC}_{OFF}-\mathrm{VC}_{ON} = x\cdot\chi_{\beta car_{S_{2}}}.
\end{align}
\end{subequations}
Here, $\mathrm{VC}_{OmO}$ is the difference between the signals recorded in the PP and PDP experiments, $x$ is the fraction of molecules removed by the dump pulse and $\chi_{j}$ the VC of species $j$. $\mathrm{VC}_{OmO}$ therefore isolates VC that is either present on or originates from $\mathrm{S}_{2}$ (FIG. \ref{fig:figC}a). 
Comparison of FIG. \ref{fig:figB}a and FIG. \ref{fig:figC}a in the 650 - 800 nm spectral region highlights the efficient removal of all ground state and solvent coherences. Two regions of clear coherent activity remain: the $\mathrm{S}_{x}$ $\leftarrow$ $\mathrm{S}_{2}$ and the $\mathrm{S}_{y}$ $\leftarrow$ $\mathrm{S}_{1}$ ESA regions with wavelength dependent coherence amplitudes reminiscent of the respective transient absorption spectra (FIG. \ref{fig:figA}b) \cite{Kumar2001,Kobayashi2008}. Short-lived VC present at early time-delays in the 650 - 800 nm window likely originates from the $\mathrm{S}_{0}$ $\leftarrow$ $\mathrm{S}_{2}$ stimulated emission transition. While the peak positions in the averaged Fourier power spectrum of the $\mathrm{S}_{1}$ region (540 - 650 nm) agree well with those obtained for an $\mathrm{S}_{1}$-only spectrum recorded via broadband impulsive vibrational spectroscopy (FIG. \ref{fig:figC}b) \cite{Liebel2013}, the intensities of the individual bands differ considerably. Both the 1538 and 1256 $\mathrm{cm}^{-1}$ modes exhibit similar intensity ratios while a number of bands in the 700 - 1300 $\mathrm{cm}^{-1}$ window, marked by arrows in FIG. \ref{fig:figC}b, are strongly enhanced in the $\mathrm{VC}_{OmO}$ spectrum.

\begin{figure}[h]
\includegraphics{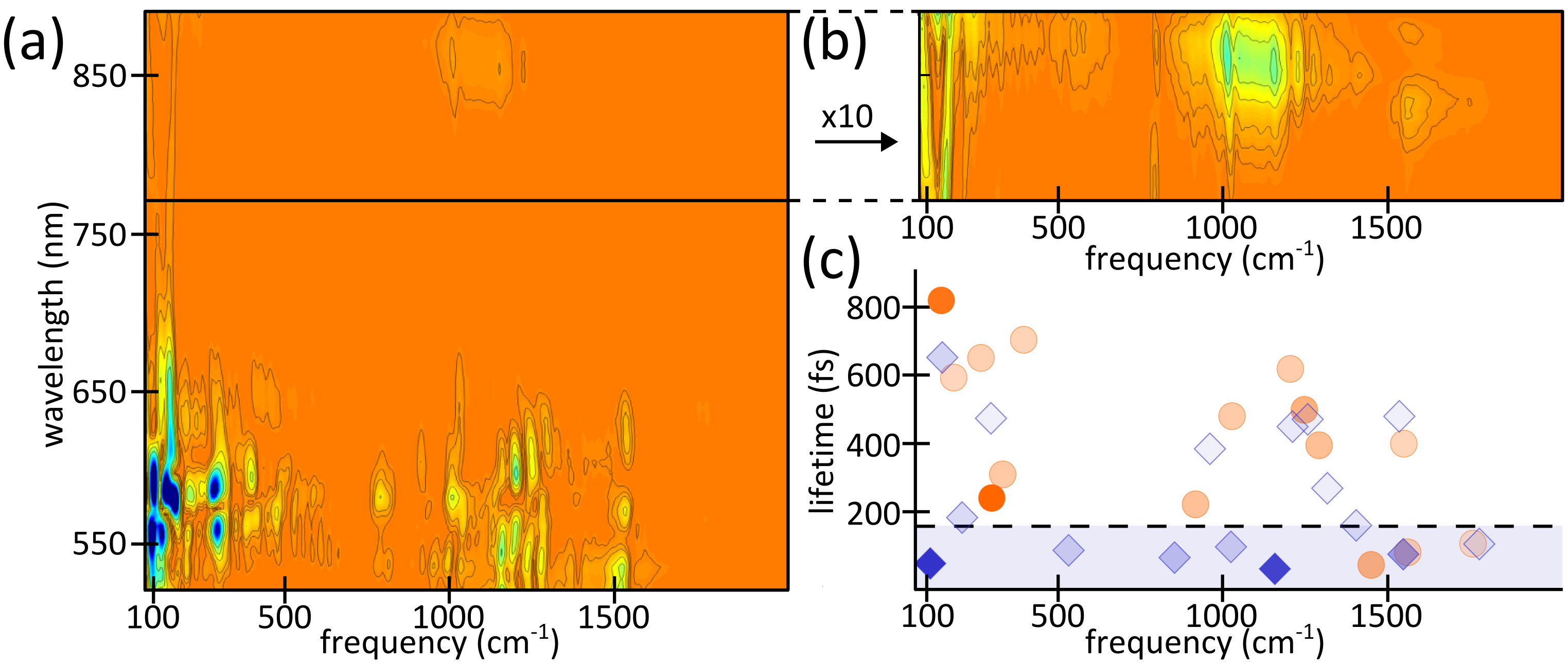}
\caption{\label{fig:figD} Fourier power maps and linear prediction singular value decomposition of the $\mathrm{VC}_{OmO}$ coherence. (a) Fourier power map as a function of detection wavelength. (b) Magnification of the $\mathrm{S}_{2}$ region. (c) Comparison of the frequencies obtained by SVD in the $\mathrm{S}_{1}$ region (orange circles) and the $\mathrm{S}_{2}$ region (blue diamonds). The symbol opacity represents the band amplitude.}
\end{figure}

In contrast to the visible, broad $\mathrm{S}_{2}$ signatures dominate the NIR (FIG. \ref{fig:figD}a,b). While the $\mathrm{S}_{2}$ VC amplitude is large (FIG. \ref{fig:figC}a), its contribution to the respective Fourier power map (FIG. \ref{fig:figD}a) is almost negligible, a result of the rapid loss of electronic population caused by the short $\mathrm{S}_{2}$ lifetime. Interestingly, magnification of the 780 - 880 nm region (FIG. \ref{fig:figD}b) reveals sharp features on top of the broad $\mathrm{S}_{2}$ bands. Fourier power spectra necessarily merge phase and amplitude information, an effect that can influence both shape and position of peaks, especially for partially overlapping bands \cite{Johnson1996}. To nevertheless obtain realistic frequency estimates and information on the vibrational dephasing times, we applied linear prediction singular value decomposition to the coherences observed in the visible and NIR regions (FIG. \ref{fig:figD}c) \cite{Johnson1996}. The majority of frequencies obtained in the $\mathrm{S}_{2}$ region exhibited short ($\textless$ 140 fs) dephasing times, in agreement with the decay time of $\mathrm{S}_{2}$. Some bands, however, such as those present in the 1100 - 1400 $\mathrm{cm}^{-1}$ region, exhibited vibrational dephasing times that clearly exceeded the electronic lifetime of $\mathrm{S}_{2}$. Interestingly, these long-lived coherences are also present in the $\mathrm{S}_{1}$ spectral region, suggesting that several $\mathrm{S}_{1}$ frequencies modulate the spectral window of the $\mathrm{S}_{2}$ ESA, even after the electronic state has nominally decayed. 

Our results clearly demonstrate that VC generated in one electronic state remains active in the product state after IC. Despite optical excitation into the $\mathrm{S}_{2}$ state with a 140 fs lifetime, VC in all known $\mathrm{S}_{1}$ modes is present throughout the $\mathrm{S}_{y}\leftarrow \mathrm{S}_{1}$ ESA window, even for vibrational periods as short as 19 fs (1785 $\mathrm{cm}^{-1}$). Therefore, the origin of the observed $\mathrm{S}_{1}$ coherence must be at time zero, but in a different electronic state. 

To understand the potential origin of the VC transfer, it is instructive to consider the role of nuclear degrees of freedom in electronic surface crossings. By definition, a conical intersection (CI) is spanned by a subset of vibrational coordinates, known as coupling modes \cite{Kuhl2002} while others, being important for reaching the CI, are classified as tuning coordinates. According to this theoretical description, nuclear wavepackets in coupling modes can be strongly influenced by passage through the CI while VC in tuning modes is largely unaffected by the surface crossing.

The spectra in FIG. 3b suggest that most of the known, intense $\mathrm{S}_{1}$ Raman bands (290, 399, 1256, 1538, 1785 $\mathrm{cm}^{-1}$) act predominantly as tuning modes. The excitation pulse generates nuclear wave packets in totally symmetric modes and this coherence is essentially unperturbed by the surface crossing. This interpretation is in line with the fundamental symmetry requirements for a $\mathrm{B}_{u}$ to $\mathrm{A}_{g}$ surface crossing requiring $\mathrm{b}_{u}$-symmetry vibrational modes to couple $\mathrm{S}_{2}$ and $\mathrm{S}_{1}$. The dramatic difference in coherent activity for a number of $\mathrm{S}_{1}$ bands after IC  (776, 1000, 1160, 1206, 1287, 1372, 1407 $\mathrm{cm}^{-1}$) compared to impulsively created VC on $\mathrm{S}_{1}$ is consistent with these modes being non-totally symmetric and therefore almost undetectable in standard Raman spectra (see Fig. 3b). In the experiments reported here the VC evident in the $\mathrm{S}_{y}$ $\leftarrow$ $\mathrm{S}_{1}$ absorption window reports on the molecule in the vicinity of the surface crossing, i.e. vibrationally hot molecules. Here, $\mathrm{S}_{1}$ and $\mathrm{S}_{2}$ are strongly mixed allowing for both totally and non-totally symmetric modes to be Raman active and thus modulate the transient electronic signals. Such appearance of non-totally symmetric vibrations has been previously reported in time-resolved anti-Stokes Raman experiments on 4-nitroaniline, although in an incoherent fashion \cite{Kozich2002}.

Given that only symmetric modes are Franck-Condon active in the $\mathrm{S}_{2}$ $\leftarrow$ $\mathrm{S}_{0}$ transition, VC in non-totally symmetric modes must therefore have been generated after excitation. The high frequencies of many detected vibrations ($\textgreater$ 1000 $\mathrm{cm}^{-1}$) together with the 140 fs lifetime of $\mathrm{S}_{2}$ firmly excludes impulsive generation of the VC by the surface crossing. Instead, the VC in non-totally symmetric modes must be generated by anharmonic coupling with the originally excited symmetric modes. Theoretically, one could differentiate between anharmonic coupling exclusively on $\mathrm{S}_{2}$ and coupling strictly induced by the CI. Given the symmetry properties of the two states ($\mathrm{B}_{u}$ and $\mathrm{A}_{g}$), however, such a differentiation appears largely semantic. The CI mixes the electronic and  symmetry characters of the two states and hence strongly distorts the potential energy surfaces of $\mathrm{S}_{2}$ and $\mathrm{S}_{1}$.

Although it is difficult to clearly assign the observed vibrations to coupling or tuning modes, we believe that the enhanced non-totally symmetric modes contribute to the coupling of the $\mathrm{S}_{2}$ and $\mathrm{S}_{1}$ and, thereby, to the formation of the CI. VC in these modes is strongly affected by the CI, although in this case it is generation of VC rather than destruction by dephasing that has been predicted if a wavepacket is already present in the coordinate prior to surface crossing \cite{Kuhl2002}. The modes are fundamentally connected to the formation of the CI because it is anharmonic coupling that generates the VC and the anharmonic coupling in turn is mediated by the CI. Non-totally symmetric modes exhibit the correct symmetry to couple the two electronic states and thereby mediate the breakdown of the Born-Oppenheimer approximation. This notion is strengthened by our observation of long-lived surface recrossings (FIG. \ref{fig:figD}) that has been predicted to occur for both coupling and tuning modes \cite{Kuhl2002}. 

The exact structural assignment of these coordinates is difficult due to the computational complexity associated with quantum chemical studies of molecules as large as $\beta$-carotene. Our methodology, however, is readily expandable to smaller molecules by using excitation pulses in the ultra-violet where structurally simpler molecules enable much more precise computational studies. The coherent nature of our methodology therefore provides scope for the elucidation of the molecular coherent dynamics associated with electronic surface crossings and thus allows for direct insight into the basics behind the breakdown of the Born-Oppenheimer approximation for polyatomic molecules in the condensed phase.

\bibliography{references}

\end{document}